\documentclass[reprint,aps,prx,amssymb,amsmath]{revtex4-2}
\usepackage{graphicx}
\usepackage{dcolumn}
\usepackage{bm}
\usepackage{caption}
\usepackage{float}  
\usepackage{amsmath}
\usepackage{physics}
\usepackage{tikz}
\usepackage{subfigure}  
\usepackage{xr-hyper} 
\usepackage{hyperref}
\makeatletter
\renewcommand{\bibitemNoStop}{}
\makeatother

\begin{document}

\preprint{APS/123-QED}

\title{Electron- and Lattice-Temperature Dependence of the Optical Response of Gold Nanoparticles}

\author{Nour E. H. Chetoui$^1$, Jonas Grumm$^{1,2}$, Robert Lemke$^2$, Andreas Knorr$^2$, and Holger Lange$^{1,}$}
\email{holger.lange@uni-potsdam.de}
\affiliation{$^1$ Institute of Physics and Astronomy, University of Potsdam, Germany\\
$^2$ Institute of Physics and Astronomy, Technical University Berlin, Germany}

\begin{abstract}
\textbf{Abstract} Transient absorption spectroscopy is routinely used to study the electron dynamics in plasmonic gold nanoparticles. Typically, the transient absorption bleach is analyzed as measure for the electron temperature. However, the implicitly assumed linear dependence between bleach intensity and temperature has not been systematically studied. Similarly, the influence of lattice heating also lacks a detailed analysis. Here, we solve momentum–resolved metal Boltzmann–Bloch equations for a semi-analytic access to the temperature-dependent gold nanoparticle absorption. We confirm the theory with steady state and transient absorption experiments, define regions of linear correlation between transient absorption bleach intensity and electron temperature and reveal a strong impact of the lattice temperature on the TA bleach intensity. 
\end{abstract}

\maketitle

\section{Introduction}
Nanoparticle (NP) plasmons,collective oscillations of the NP electron gas with the field of light, condense optical energy into small volumes below the wavelength. The decay dynamics of the excited electronic states proceeds through a series of events.
Orientational relaxation scatters the momentum--polarized electrons into unpolarized, isotropic non--thermal electrons which thermalize via electron--electron and electron--phonon interaction, resulting in a hot thermalized electron gas \cite{grumm_femtosecond_2025,khurgin_hot_2024,wach_dynamics_2025}. Electron--phonon interaction cools the electrons gas accompanied by a simultaneous heating of the ion-lattice\cite{qiu_heat_1993,tzou_macro-_1997}.
\\
A vast body of research investigates the details of the plasmon dynamics in fundamental solid state physics and in context of applications such as photocatalysis\cite{brongersma_plasmon-induced_2015,jain_plasmonic_2022}.
The most direct access to the plasmon dynamics is via the absorption of the thermalized electron gas.
In transient absorption (TA) pump--probe experiments with metal NPs, an optical pump pulse excites a plasmon. The subsequent relaxation increases the electron temperature which leads to a change of the metal's dielectric function and a broadening of the plasmon resonance, detected as bleach in the white light probe pulse absorption.
Typically, the time-resolved TA bleach is fitted by exponential functions and discussed in two--temperature models (electrons and lattice)\cite{hartland_optical_2011}.
This implies that the TA bleach intensity (reduction of absorption) is proportional to the electron temperature. A broad range of works discuss the underlying dependency of the metal's complex dielectric function on the electron temperature \cite{schirato_ultrafast_2023,okeeffe_disentangling_2021,brown_experimental_2017}.
However, a comprehensive reasoning for the linear relation of electron temperature and TA bleach intensity is still missing. In addition, the common calculation of the electron cooling time from the TA bleach dynamics is only a valid approach for situations where lattice temperature changes constitute no relevant source of TA contrast.

This is the case for the early few picoseconds of the electron dynamics after excitation, where changes of the electron temperature are much more drastic then the heating of the lattice.
At longer times, a NP's lattice can easily be heated by tens of Kelvins and the elevated lattice temperature can also impact the NP absorption. In such a scenario, the TA bleach intensity is not necessarily a direct measure for the electron temperature, yielding related lattice heating and heat dissipation discussions based on two-- and three--temperature models questionable.\\
In this work, we set up TA experiments and develop momentum--resolved metal Boltzmann--Bloch equations providing the frame for the calculation of the gold NP (AuNP) optical response under equilibrium and non--equilibrium conditions. The applied theory based on a microscopic evaluation fo the electron-phonon scattering rates, reproduces the experimental spectra and explains the observed temperature dependencies of the AuNP absorption.
For scenarios with a temperature gradient between electrons and lattice, we reveal a strong impact of the lattice temperature on the AuNP absorption. For fixed lattice temperatures, the intensity--temperature relation is confirmed to be quasi--linear in most regimes, while the total absorption intensity is strongly impacted by lattice temperature changes. This is particularly relevant for  or in extreme pump scenarios and for long delays, for example when discussing thermal dissipation.

\section{Theory Framework} \label{sec:theory}
Our theoretical approach assumes that after a thermalization time of a few hundred femtoseconds \cite{fann_direct_1992, mueller_relaxation_2013}, phonons and pump-induced hot electrons are distributed thermally in a quasi--equilibrium. In this state, temperatures are defined for electrons ($T_e$) and phonons ($T_l$) with $T_e \neq T_l$. The linear optical response is then calculated for this quasi--equilibrium situation:

The TA bleach relates the NP extinction cross-section (absorbance) $C_{ext}$ to a chosen reference, $TA = C_{ext} - C_{ext}^{ref}$. In quasi--static approximation, the absorbance of the NP is given by \cite{bohren_absorption_2008}
\begin{align}
    C_{ext}(\omega) =& \frac{\omega}{c} \Im{\alpha(\omega)} + \frac{\omega^4}{6\pi c^4} |\alpha(\omega)|^2 ~, \label{eq:extinction}
\end{align}
where the first term accounts for the absorption cross-section, which typically dominates the absorbance, and the second term reflects the scattering cross-section.
The polarizability $\alpha(\omega)$ occurring in both contributions is given for a spherical AuNP in the dipole approximation of Mie theory \cite{bohren_absorption_2008} by
\begin{align}
    \alpha(\omega) = 4\pi R^3 \frac{\varepsilon_{Au}(\omega) - \varepsilon_{out}}{2\varepsilon_{out} + \varepsilon_{Au}(\omega)} \label{eq:polarizability}
\end{align}
with NP radius $R$, permittivity of the surrounding medium $\varepsilon_{out}$, and gold bulk dielectric function
\begin{align}
    \varepsilon_{Au}(\omega) = \varepsilon_b + \chi_{intra}(\omega, T_e, T_l) + \chi_{inter}(\omega, T_e, T_l)  \label{eq:AU_dielectric_function}
\end{align}
with static background permittivity $\varepsilon_b$ caused by all occupied electrons not treated resonantly. Here, $\varepsilon_b$, fixed by the stationary absorption experiments, will serve as the only fit parameter. In Eq.~\eqref{eq:polarizability}, the plasmon is formed by confining the quasi--free conduction band electron gas with intraband susceptibility $\chi_{intra}(\omega)$ into the NP due to Maxwell boundary conditions. The hybridization of the plasmon with interband transitions included in $\chi_{inter}(\omega)$ in Eq.~\eqref{eq:polarizability} additionally shifts and broadens the plasmon resonance \cite{khurgin_hot_2024, zoric_gold_2011, pirzadeh_plasmon_2014, tserkezis_self-hybridisation_2024}, so that a comprehensive theory of the temperature--dependent NP plasmon must address both intra-- and interband processes.
\\
The temperature dependencies of the susceptibilities $\chi_{intra/inter}$ are obtained from a comprehensive analysis of electron--phonon scattering on the microscopic scale: In Sec. IV A of the Supplemental Material we summarize how the susceptibilities are derived -- without any fit parameters to the TA experiments -- based on a microscopic formulation of linearized metal Boltzmann--Bloch equations incorporating electron--phonon scattering \cite{lemke_boltzmann-bloch_2026, haug_quantum_2009}. 
\\

We can measure the electronic transitions in the quasi-free conduction band by measuring the optical response (absorbance) of the hot thermalized electron gas, Eq.~\eqref{eq:chi_intra}, as given by the Drude susceptibility
\begin{figure}
    \centering
    \includegraphics[width=1\linewidth]{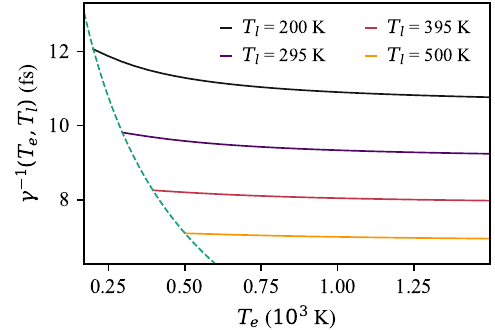}
    \caption{Inverse two-temperature relaxation rate $\gamma^{-1}(T_e,T_l)=(\gamma^{(ep)}_{k_F}(T_e,T_l) + \gamma_S)^{-1}$ for selected phonon temperatures $T_l$. The dashed line represents the rate at $T_e=T_l$.}
    \label{fig:rta}
\end{figure}
\begin{align}
    \chi_{intra}(\omega, T_e, T_l) = - \frac{\omega_p^2}{\omega^2 + i \omega \gamma(T_e, T_l)} \label{eq:chi_intra}
\end{align}
with bulk plasma frequency $\omega_p$ and relaxation rate $\gamma(T_e, T_l)$. Interband transitions between the highest $5d$ valence band ($v$) and the $6sp$ conduction band ($c$) in gold are addressed by \cite{lemke_boltzmann-bloch_2026}
\begin{align}
    &\chi_{inter}(\omega, T_e, T_l) \label{eq:chi_inter} \\
    &= \sum_\mathbf{k} \frac{A_\mathbf{k}^{vc} (f^{FD}(\epsilon_\mathbf{k}^c, T_e) - f^{FD}(\epsilon_\mathbf{k}^v, T_e))}{\omega^2 - \frac{(\epsilon_\mathbf{k}^c -\epsilon_\mathbf{k}^v)^2}{\hbar^2} - \zeta_\mathbf{k}^2(T_e, T_l) + 2i \omega \zeta_\mathbf{k}(T_e, T_l)} ~, \nonumber
\end{align}
where $\epsilon_\mathbf{k}^\lambda$ are the electronic energies of the bands $\lambda=v,c$ with momentum $\vb{k}$ modeled as in \cite{rosei_d_1973}, $f^{FD}(\epsilon_\mathbf{k}^\lambda, T_e)$ are the Fermi-Dirac distributed occupations, $A_{\vb{k}}^{vc}$ is an amplitude factor (see Sec.~IV A of Supplemental Material), and $\zeta_\mathbf{k}(T_e, T_l)$ is the dephasing rate. As a benchmark, the calculated gold bulk dielectric function $\varepsilon_{Au}$ agrees with typical experimental data \cite{johnson_optical_1972, lemke_boltzmann-bloch_2026} at room temperature. All parameters used are listed in Sec.~IV Tab.~I in the Supplemental Information.
\\
The linearized intraband electron--phonon scattering in the conduction band yields the two--temperature dependent orientational relaxation rate 
\begin{subequations}
    \label{eq:2T_or_rate}
     \begin{align}
        \gamma^{(ep)}_k =& \gamma^{(ep),N}_k + \gamma^{(ep),U}_k ~, \\
        \gamma^{(ep), N}_{k} =& \frac{V m^c}{4\pi \hbar^3 k_B T_e} \int_0^{q_D} \mathrm{d}q~ \frac{q^3}{k^3}  \Phi_q(T_e,T_l) \\
        &\times |g_{q,0}|^2 \Big( \Sigma^+_q(T_e,T_l) + \frac{\hbar \omega_q}{\epsilon_q^c} \Sigma^-_q(T_e,T_l) \Big)~, \nonumber \\
        \gamma^{(ep), U}_{k} =& \frac{V m^c}{8 \pi \hbar^3 k_B T_e} \sum_{\vb{G} \neq 0} \int_{G-2k_F}^{q_D} \mathrm{d}q~ \frac{q^3}{Gk^3} \Phi_q(T_e,T_l) \\
        \times \int_{G-q}^{2k_F} &\mathrm{d}p~|g_{q,p-q}|^2 \Big( \frac{p^2}{q^2} \Sigma^+_q(T_e,T_l) + \frac{\hbar \omega_q}{\epsilon_q^c} \Sigma^-_q(T_e,T_l) \Big) ~, \nonumber
    \end{align}   
\end{subequations}
where the terms $N$ and $U$ distinguish between normal and Umklapp electron--phonon scattering processes with reciprocal lattice vectors $\vb{G}$ \cite{pines_elementary_2018}. Here, $m^\lambda$ is the effective electron mass, $V$ the volume, $k_F$ the Fermi momentum, $q_D$ the Debye wave number and $g_{\vb{q,G}}$ the electron--phonon matrix element (see Sec. IV A of Supplemental Material).
The temperature dependence is included in the factors
\begin{align}
    \Phi_q(T_e, T_l) =& -\hbar \omega_q n(\hbar \omega_q, T_l) n(-\hbar \omega_q, T_e) ~, \\
    \Sigma^\pm_q(T_e,T_l) =& 1 \pm \exp(\hbar \omega_q \Big( \frac{1}{k_B T_l} - \frac{1}{k_B T_e} \Big)) \label{eq:def_sigma}
\end{align}
with Bose--Einstein distribution $n(\hbar \omega_q,T)=(\exp(\hbar \omega_q/k_BT)-1)^{-1}$ and dispersion of longitudinal acoustic phonons $\hbar \omega_q$.
In the limit of equilibrated temperatures $T_e=T_l$, the theory delivers the linear optical response of the steady state and Eq.~\eqref{eq:2T_or_rate} reproduces the well-known Bloch--Grüneisen formula \cite{bloch_zum_1930, lawrence_umklapp_1972, grumm_theory_2025}. Beyond that, the extension to the case $T_e \neq T_l$ allows us to study quasi--equilibrium systems without a numerically challenging evaluation of the full scattering dynamics \cite{grumm_theory_2025}. The total relaxation rate 
\begin{align}
    \gamma(T_e,T_l) = \gamma^{(ep)}_{k_F}(T_e,T_l) + \gamma_S
\end{align}
in Eq.~\eqref{eq:chi_intra} is obtained by evaluating Eq.~\eqref{eq:2T_or_rate} at the Fermi momentum \cite{grumm_femtosecond_2025} and considering in addition surface--assisted Landau damping in AuNP \cite{uskov_broadening_2014, shahbazyan_landau_2016}
\begin{align}
    \gamma_{S} = \frac{3 \hbar k_F}{4Rm^c} 
\end{align}
introduced here on a phenomenological basis as temperature--independent parameter.
\\
For the dephasing of the interband transitions in Eq.~\eqref{eq:chi_inter}, we find in a dephasing rate approximation \cite{hess_maxwell-bloch_1996} the two--temperature dephasing rate 
\begin{subequations}
    \label{eq:dra_rates}
     \begin{align}
        \zeta_k =& \zeta^{(ep),N}_k + \zeta^{(ep),U}_k ~, \\
        \zeta^{(ep), N}_{k} =& \frac{V}{4\pi \hbar^3 k_B T_e} \sum_{\lambda = v,c} m^\lambda \int_0^{q_D} \mathrm{d}q~\frac{q}{k} \\
        &\times \Phi_q(T_e, T_l) |g_{q,0}|^2 \Sigma^+_q(T_e,T_l) ~, \nonumber \\
        \zeta^{(ep), U}_{k} =& \frac{V}{8\pi \hbar^3 k_B T_e} \sum_{\lambda = v,c} m^\lambda \sum_{\vb{G} \neq 0} \int_{G-2k_F}^{q_D} \mathrm{d}q~ \frac{q}{Gk} \\
        &\times \Phi_q(T_e, T_l) \int_{G-q}^{2k_F} \mathrm{d}p |g_{q,p-q}|^2 \Sigma^+_q(T_e,T_l) ~, \nonumber
    \end{align}   
\end{subequations}
which provides a similar temperature-dependency as in the intraband relaxation rate in Eq.~\eqref{eq:2T_or_rate}.

In a simplified picture, the temperatures of electrons and phonons are reflected in the TA bleach intensity as follows: The plasmon resonance is approximated in lowest order by neglecting the temperature and frequency dependency of the interband susceptibility so that the gold dielectric function in Eq.~\eqref{eq:polarizability} becomes $\varepsilon_{Au}(\omega,T_e,T_l) \approx \Tilde{\varepsilon}_b + \chi_{intra}(\omega,T_e,T_l)$ with effective background $\Tilde{\varepsilon}_b$. Then, evaluating in Eq.~\eqref{eq:extinction} at the plasmon resonance $\omega_0 = \omega_p/\sqrt{2\varepsilon_{out}+\Tilde{\varepsilon}_b}$, the absorbance shows a proportionality to the inverse total relaxation rate $\gamma$
\begin{align}
    C_{ext}(\omega_0,T_e, T_l) \sim \frac{1}{\gamma(T_e,T_l)} = \frac{1}{\gamma^{(ep)}_{k_F}(T_e,T_l) + \gamma_S} ~. \label{eq:ext_at_LSP}
\end{align}
Thus, the TA bleach intensity at the plasmon resonance directly reflects the joint changes of electron and phonon temperature via the orientational relaxation rate $\gamma_{k_F}^{(ep)}(T_e,T_l)$. Fig.~\ref{fig:rta} displays the inverse two-temperature relaxation rate $\gamma^{-1}(T_e,T_l)$ as timescale of the plasmon decay. Fig.~\ref{fig:rta} suggests that the influence of electron temperature is highest at low phonon temperatures. This mostly originates from the factor $\Sigma_q^+$, Eq.~\eqref{eq:def_sigma}. However, the strongest contrast in the TA bleach is expected to be related to changing phonon temperatures. We will disentangle the $T_e$ and $T_l$ signals that comprise the TA bleach intensity in the following discussion of the experimental results.
\section{Results and discussion}
\subsection{Steady--state experiment}
\begin{figure}[htb]
    \centering
    \includegraphics{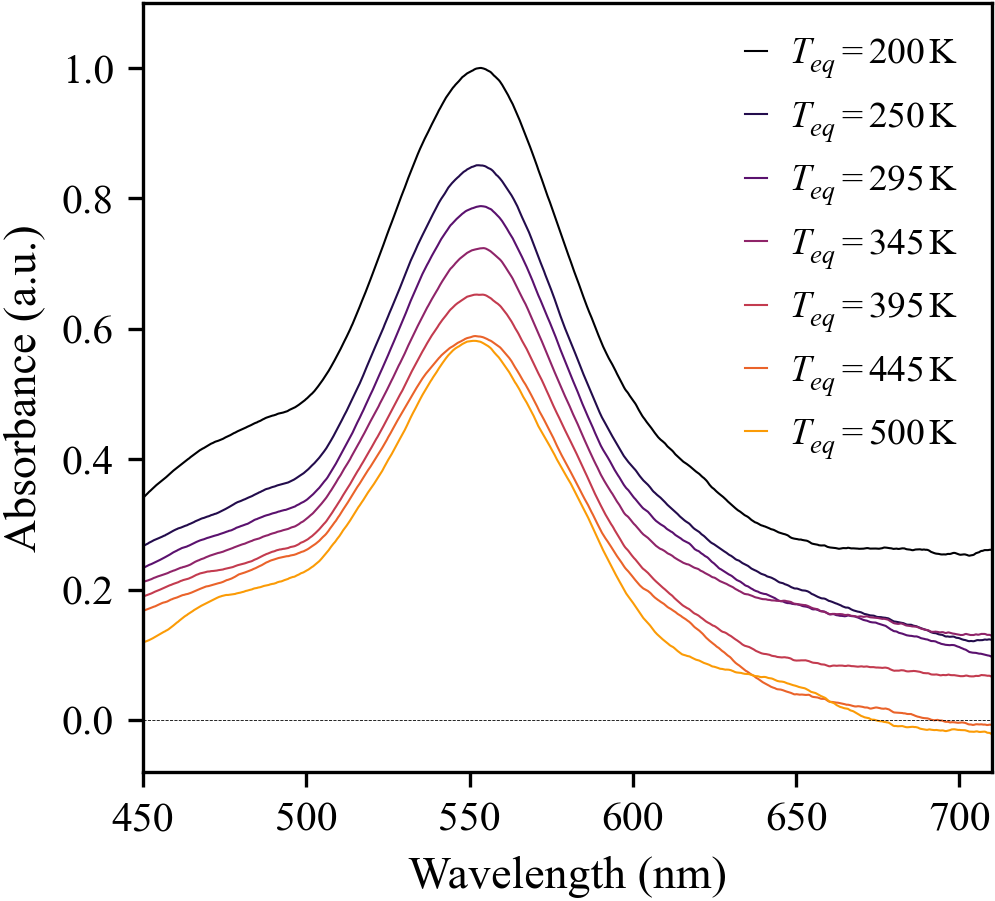}
    \caption{Temperature--dependent absorption spectra of AuNP. The spectra were normalized such that the relative ratios are preserved.}
    \label{fig:cw-abs}
\end{figure}

Fig.~\ref{fig:cw-abs} displays temperature--dependent stationary steady state absorption spectra of 40~nm diameter AuNPs with the temperature set by a cryostat.
\begin{figure*}
  \includegraphics{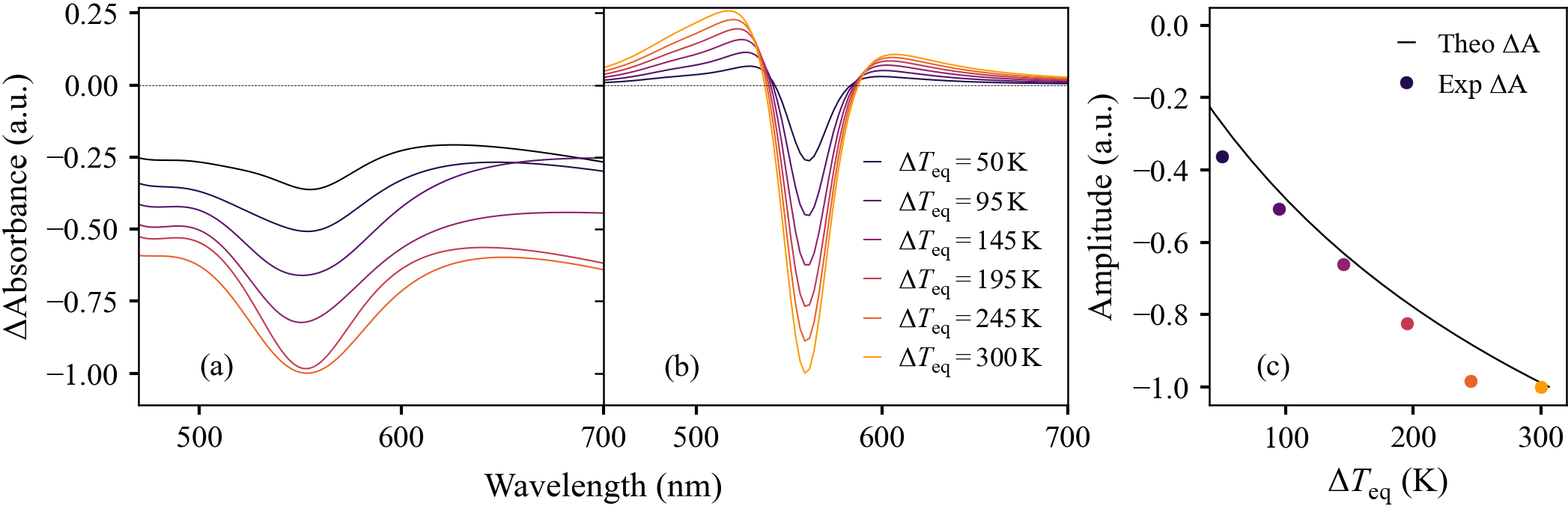}
    \caption{(a) Differential steady--state absorption spectra from the data plotted in Fig.~\ref{fig:cw-abs}. The spectra are referenced to the $T_{eq}=200$~K spectrum. (b) Calculated differential absorption spectra for the same temperatures as for (a). The spectra in (a) and (b) are scaled by a factor defined by the absolute value of the most intense $\Delta A$. The dashed lines represent the baseline of the $200$~K reference.
    (c) Experimental and calculated differential absorption bleach versus equilibrium temperature increase $\Delta T_{eq}$ relative to $200$~K. Note that the theory curve consists of individual data points and is not a fit to the experimental values.}
    \label{fig:cw-exptheo}
\end{figure*}
In this case, the electrons and the lattice are in thermal equilibrium and share a joint equilibrium temperature $T_e=T_l=T_{eq}$.
With increasing $T_{eq}$, the plasmon absorption peak around $\lambda_0 \sim 550$ nm slightly redshifts, broadens, and decreases in amplitude. This originates mainly from the temperature dependence of the gold dielectric function, Eq.~\eqref{eq:AU_dielectric_function}: heating increases the electron--phonon scattering and the relaxation rate $\gamma$, leading to a larger imaginary part of the dielectric function.
This results in a broadening of the plasmon linewidth and a reduced peak magnitude of the absorbance as in Eq.~\eqref{eq:ext_at_LSP}. Simultaneously, a temperature--induced change in the real part of the dielectric function shifts the resonance condition toward lower energies, resulting in a redshift \cite{yeshchenko_temperature_2013}.\\
As most electron dynamics studies of AuNPs are based on TA,
differential $\Delta T_{eq}$--dependent steady-state absorption spectra versus the lowest temperature (200~K) are displayed in Fig.~\ref{fig:cw-exptheo}(a).

The $\Delta T_{eq}$--dependent change of the AuNP absorption leads to an absorption bleach centered around the plasmon absorption. The corresponding theory data is plotted in Fig.~\ref{fig:cw-exptheo}(b). Fig.~\ref{fig:cw-exptheo}(c) displays the bleach amplitude versus $\Delta T_{eq}$. Our theoretical framework, based on momentum--resolved metal Boltzmann-Bloch equations, reproduces these trends.
In principle, the distinct temperature-amplitude relation enables assessing AuNP temperature changes from straightforward (transient) absorption measurements.% \andreas{The different zero lines are not explained in Fig. 3a,b!} \nour{added in the caption.}

\subsection{Transient absorption experiment}
Directly after optical excitation,
the electron system is in a nonthermal state and reaches a thermal quasi--equilibrium on sub--picosecond timescales with elevated temperature $T_e > T_l$. The heat dissipation from the excited hot electron gas to the lattice occurs in a picosecond time range.
\subsubsection{Long delays}
\begin{figure}
    \includegraphics{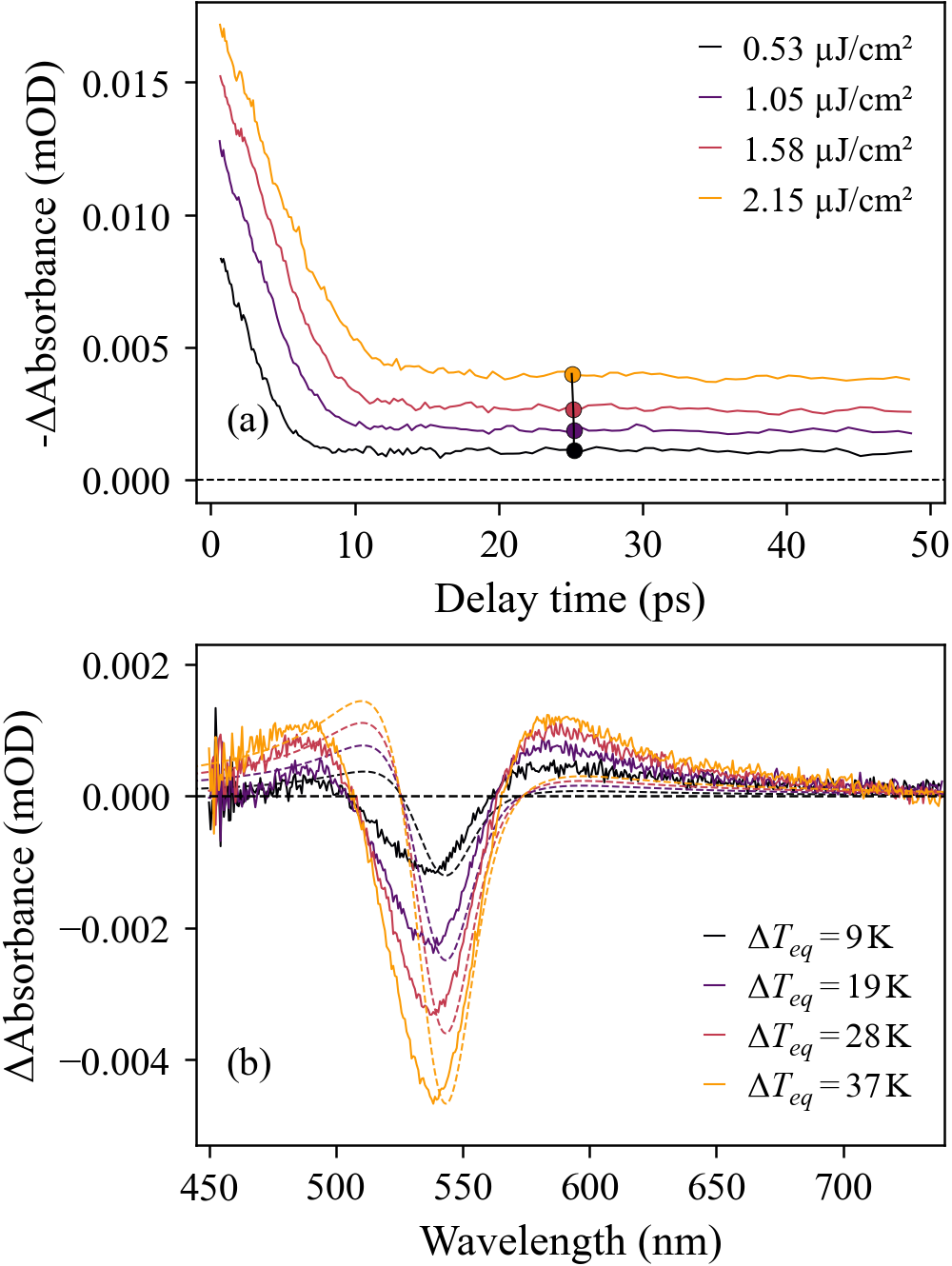}
    \caption{(a) Pump–probe TA bleach dynamics at $T_{eq,0} = 295$~K for different pump fluences. \newline
    (b) Experimental \textbf{(${-}$)} and theoratical \textbf{(${--}$)} TA spectra at $25$~ps delay, as highlighted in (a). The calculated spectra are scaled such that the bleach intensities at $\Delta T_{eq}= 37$~K match.}
    \label{fig:longdelays}
\end{figure}
\begin{figure*}
    \includegraphics{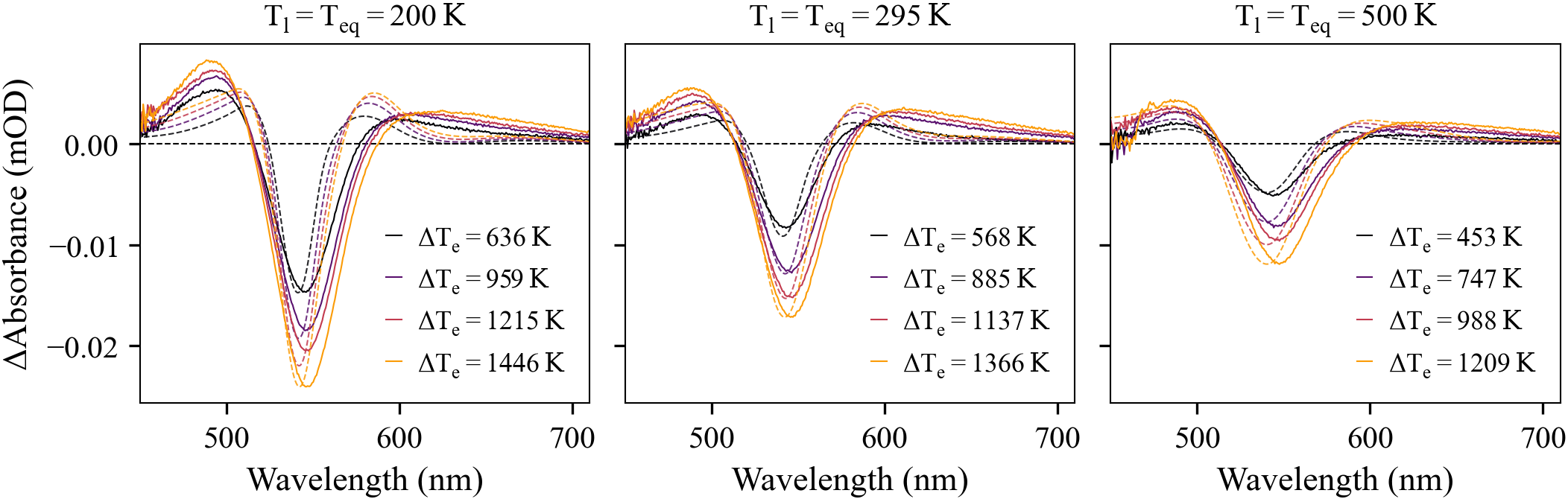}
    \caption{Transient absorption spectra recorded at the delay time of maximum bleach contrast for equilibrium lattice temperatures $T_l = T_{eq}$. At each temperature, spectra are shown for four excitation energies labeled by the corresponding initial electron temperature rise $\Delta T_e$ (i.e., different pump fluence). Dashed curves show calculated TA referenced to the equilibrium spectrum and scaled so that the bleach minimum matches the experimental minimum.}
    \label{fig:comp}
\end{figure*}
At longer time scales after optical excitation ($>$10~ps, cf. Fig.~\ref{fig:longdelays}a), the electrons and the lattice reach thermal equilibrium at an elevated temperature with $T_e=T_l=T_{eq} > T_{eq,0}$ and $T_{eq,0}$ the initial temperature set by the cryostat. As heat dissipation into the environment occurs on much longer timescales the system can be considered in a quasi--stationary state with an excitation--dependent increase in equilibrium temperature $\Delta T_{eq}$
(cf. Fig.~\ref{fig:longdelays}(a) and Supplemental Material). Fig.~\ref{fig:longdelays}(b) demonstrates this quasi--stationary state starting from the $25$~ps time delay mark, each corresponding to the increase in $T_{eq}$ calculated based on the absorbed energy per AuNP and the electron and lattice heat capacities (see Sec. III B of Supplemental Material). The absorption change is reproduced by the theory, reaffirming its validity for equilibrium conditions. For long delays ($> 25$ ps), the TA bleach is thus a direct measure for AuNP temperature changes. As AuNP cooling occurs on  timescales where electrons and lattice share an equilibrium temperature, heat dissipation can be quantitatively followed with TA experiments.

\subsubsection{Short delays}
\begin{figure}[htb]
     \includegraphics{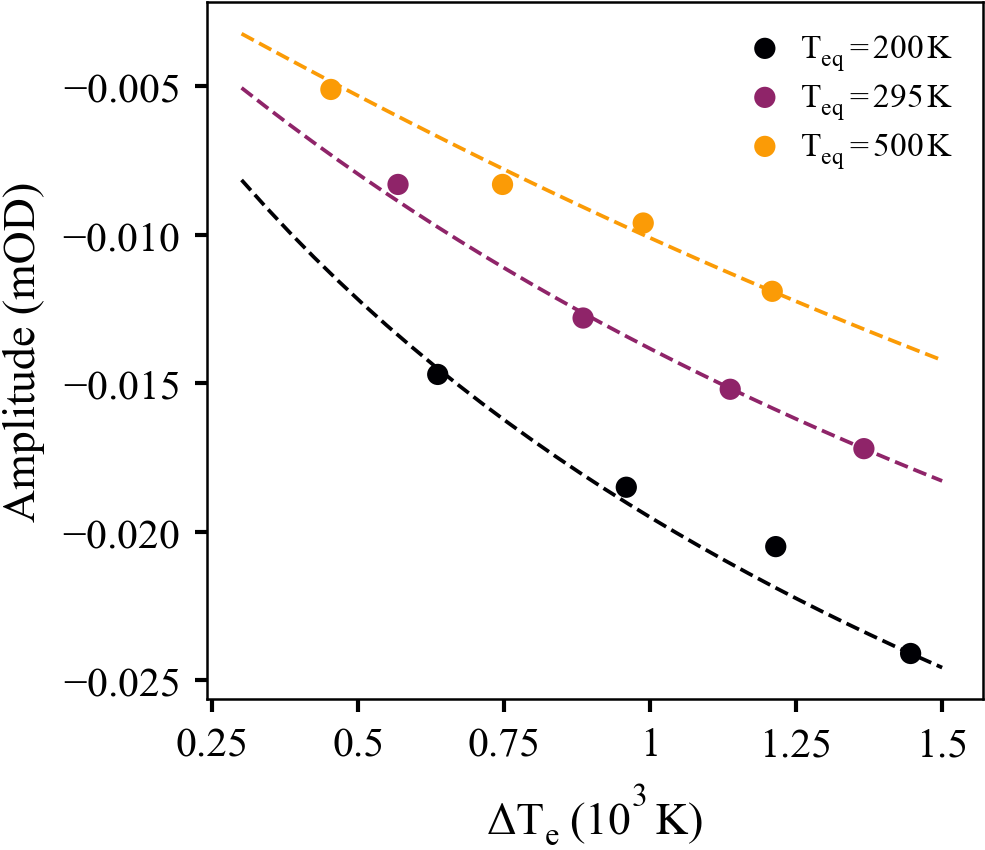}
     \caption{TA bleach intensity at different $T_l = T_{eq}$ as a function of the initial electron temperature rise $\Delta T_e$.
     $\bullet$: bleach intensities extracted from the TA spectra. $--$: calculated bleach intensities, scaled by a constant value to match the experimental amplitude. Note that the theory curve is not a fit to the experimental values.}
     \label{fig:d}
 \end{figure}

Shortly after optical excitation ($\sim 500-600$~fs) and after electron thermalization, the TA bleach intensity at its maximum.
At this point in time, one can assume negligible energy transfer to the lattice with $T_l=T_{eq}$.
Fig.~\ref{fig:comp} displays TA spectra for different cryostat ($T_{eq}=T_l$) temperatures at their maximum contrast each for an assortment of initial electron temperature gradients $\Delta T_e = T_e-T_{eq}$,
It is apparent that the TA bleach intensity is no direct measure for the change in electron temperature $\Delta T_e$.
For example, The data sets for $T_l=200$~K and $T_l=500$ both each feature a spectrum for an electron temperature gradient of $\Delta T_e \approx 1200$~K and very different absorbance changes.
The TA bleach intensity strongly depends on $T_l=T_{eq}$. In addition to the experimental TA spectra, Fig.~\ref{fig:comp} displays calculated TA spectra with the two temperatures $T_e$ and $T_l$ as external parameters. The agreement of the TA bleach intensity is very good, confirming the validity of the two--temperature dependent theory and the necessity of treating electrons and lattice as separate subsystems with individual temperatures $T_e$ and $T_l$ to reproduce the experimental observations, solving problems of earlier attempts \cite{ferrera_thermometric_2020}.
Figure~\ref{fig:d} displays the TA bleach intensities across different $\Delta T_e$ values, alongside the corresponding theoretical data. 
 
For fixed lattice temperatures, the dependence of the TA bleach intensity on $\Delta T_e$ is mostly linear at high excitation levels and high lattice temperatures, but deviates progressively as $\Delta T_e$ decreases, with the deviation being most pronounced at low $T_l$. This behavior originates from the fact that the nanoparticle absorbance at plasmon resonance is proportional to the inverse relaxation rate $\gamma^{-1}(T_e,T_l)$, Eq.~\eqref{eq:ext_at_LSP}, which varies nonlinearly with temperature and is not directly reflected in the optical response. At low pump intensities, the nonlinear relation emerges from two effects also appearing in Fig.~\ref{fig:rta}: small changes in $\Delta T_e$ produce larger variations in the relaxation rate at low $T_{eq}$, independent from $T_e$, while rising $T_{l}$ increasingly modulates the bleach signal.

Deviations of the calculated TA from the experiment could originate from electron-electron scattering neglected in the theory: In quasi--stationary situations, the electron-electron contribution to orientational relaxation is typically negligible \cite{lawrence_electron-electron_1973}. However, in quasi--equilibrium with $T_e \gg T_l$, low lattice temperatures reduce electron-phonon scattering (cf.~Fig.~\ref{fig:rta}) while pump--induced elevated electron temperatures increase electron-electron scattering which depends only on $T_e$ and thus results in deviations from the theoretical TA spectrum. 

\subsubsection{Intermediate delays (1-10~ps)}
\begin{figure*}
    \includegraphics{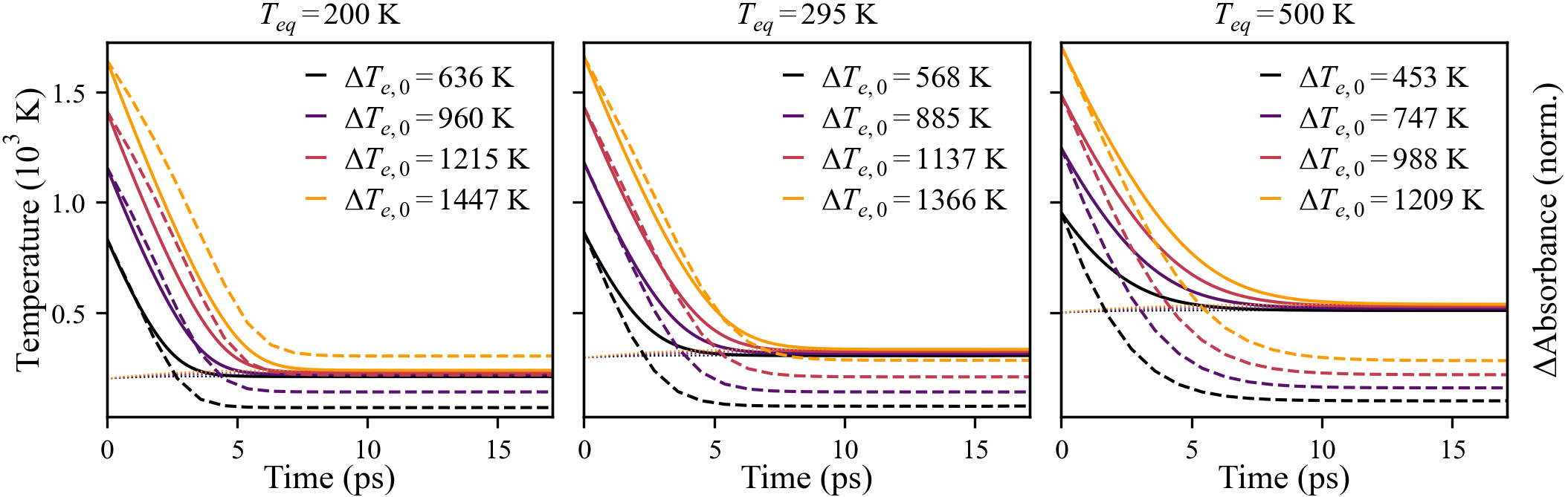}
    \caption{Calculated AuNP TA bleach intensity dynamics for three $T_{eq}$ and a set of $\Delta T_{e,0}$ in the range of common initial temperature rises in TA experiments (cf. Fig.~\ref{fig:d}). The dynamics of $T_e$ and $T_l$ are calculated from a two-temperature model with typical AuNP electron-phonon coupling constant \cite{staechelin_size-dependent_2021}. Solid curves correspond to $T_e$, dotted curves to $T_l$, and dashed curves to the TA bleach intensity.}
    \label{fig:TTM}
\end{figure*}

As electron and lattice heat capacities are very different, the level of deviation from a linear relation between $T_e$ and the intensity of the TA bleach will vary with the point in time of observation and with the amount of deposited energy. To illustrate such effects, Fig.~\ref{fig:TTM} displays evaluations of a two--temperature model for different $T_l=T_{eq}$ and $\Delta T_{e,0}$ as initial conditions (see Sec. IV B of Supplemental Material) and compares the temperatures to the calculated TA bleach intensity.

For the early dynamics ($<5$~ps), $\Delta T_e$ and the TA bleach intensity mostly share a linear dependency. For longer times ($>10$~ps), the TA bleach is dominated by the absorption change originating from the joint $T_l=T_e$. In intermediate times, the TA bleach intensity does not fully follow $\Delta T_e$. For small $T_{eq}$, large pump intensities lead to a strong deviation from the linear dependency of $\Delta T_e$ and TA bleach intensity. For room temperature calculations, the discrepancy is more pronounced for low pump intensities and even more for high $T_{eq}$. Thus, for conditions such as low pump intensity or high initial temperature, the direct extraction of electron cooling times from TA dynamic traces is questionable. Reported values from older studies are probably defective.
To circumvent such problems, the early dynamics of AuNPs excited with modest fluences are best investigated. For long times, the TA bleach intensity needs to be interpreted as measure for the quasi--stationary join temperatures $T_l=T_e$ and intermediate times always require a careful discussion.

\section{Conclusions}
In this study, we investigated how the optical response of plasmonic AuNPs depends on the electron and lattice temperatures under both equilibrium and quasi--equilibrium conditions. We combined temperature--dependent steady--state absorption with pump--probe transient absorption (TA) measurements and developed a framework based on momentum--resolved metal Boltzmann--Bloch equations. Within this framework, the measured absorption and TA spectra are reproduced across temperature and fluence, and the observed trends are rationalized by the temperature dependence of electron--phonon scattering and its impact on the dielectric function, which governs the plasmon resonance response.
The bleach intensity is not uniquely determined by the initial electron temperature change, the optical response is strongly modulated by the lattice temperature.
For common TA scenarios, the commonly used assumption---namely, that TA bleach intensity (or a single--wavelength kinetic trace) can be treated as a direct proxy for the electron temperature change is valid. However, the interplay of lattice and electron temperature implies that the procedure is generally unreliable once lattice heating becomes appreciable.
Accurate investigation of ultrafast electron dynamics therefore requires interpreting TA spectra with a model that includes both $T_e$ and $T_l$, rather than assuming a universally linear mapping between bleach amplitude and electron temperature. Practically, this motivates focusing on the earliest dynamics (typically $<5$~ps) under sufficiently high excitation, where the signal is most sensitive to $T_e$ and least contaminated by concurrent lattice heating, while treating long--delay TA signals primarily as a thermometer for the equilibrated temperature $T_{eq}$.

\section{Methods}
\textbf{Samples and Temperature Control}\\
AuNPs with a diameter of $40$ nm were synthesized using a seeded growth approach \cite{schulz_optimizing_2022}. The colloidal solution was spin--coated onto glass substrates and mounted within a  cryostat (Oxford CryostatN).\\
\textbf{Steady-State Optical Characterization}\\
Temperature-dependent stationary absorption spectra were acquired across a temperature range of $200$ to $500$ K.\\
\textbf{Transient Absorption Spectroscopy:} TA measurements were performed using a Ti:Sapphire-based laser system ($800$ nm, doubled to 400~nm, $35$ fs pulse duration, $1$ kHz repetition rate, Spectra Physics) and a commercial TA setup (Ultrafast Systems HELIOS). The cryostat was integrated into the TA setup for $T_{eq}$ control. To disentangle the interdependent effects of electron and lattice heating, data were analyzed across two primary temporal regimes: short-delay dynamics ($<5$~ps) and long-delay dynamics ($25$~ps).
\section*{Acknowledgments}
J.G. and H.L. acknowledge funding by the Deutsche Forschungsgemeinschaft (DFG, German Research Foundation) -- CRC/SFB 1636 -- Project ID 510943930 -- Project No. A08 and Project ID 504656879. We thank Dominik Hoeing and Yannic Staechelin for support with the cryogenic measurements, Felix Stete and Wouter Koopman for fruitful discussions.

\section*{Data availability}
The data that supports the findings of this article are not publicly available upon publication. The data are available from the authors upon reasonable request.

\section*{Supplemental Material}
The Supplemental Material at [URL] contains details on sample preparation, experimental methodology, theoretical background, additional comparisons between experiment and theory and additional references \cite{schulz_optimizing_2022,kaganov_relaxation_1957,anisimov_electron_1974,allen_theory_1987,ortolani_pump-probe_2019,aruda_identification_2013,cordoba_heat-capacity_1971,ashcroft_solid_1976}.

\end{document}